# Why Johnny can't rely on anti-phishing educational interventions to protect himself against contemporary phishing attacks?

**Matheesha Fernando**
Department of Computer Science and Information Technology
School of Engineering and Mathematical Sciences
La Trobe University
Melbourne, Australia
m.fernando@latrobe.edu.au

**Nalin Asanka Gamagedara Arachchilage**
Department of Computer Science and Information Technology
School of Engineering and Mathematical Sciences
La Trobe University
Melbourne, Australia
nalin.asanka@latrobe.edu.au

## Abstract

Phishing is a way of stealing people's sensitive information such as username, password and banking details by disguising as a legitimate entity (i.e. email, website). Anti-phishing education considered to be vital in strengthening "human", the weakest link in information security. Previous research in anti-phishing education focuses on improving educational interventions to better interact the end user. However, one can argue that existing anti-phishing educational interventions are limited in success due to their outdated teaching content incorporated. Furthermore, teaching outdated anti-phishing techniques might not help combat contemporary phishing attacks. Therefore, this research focuses on investigating the obfuscation techniques of phishing URLs used in anti-phishing education against the contemporary phishing attacks reported in PhishTank.com. Our results showed that URL obfuscation with IP address has become insignificant and it revealed two emerging URL obfuscation techniques, that attackers use lately, haven't been incorporated into existing anti-phishing educational interventions.

# 1   Introduction

With the advancement of the Internet, World-Wide-Web and mobile devices, ordinary people are increasingly adapting to rely on technology in every aspect of their modern life ranging from online shopping, socialising, education to entertainment (Arachchilage et al., 2016). As people's reliance on the internet grows, Internet fraud and cybercrime have become a greater threat for both individuals and organisations (Arachchilage and Love, 2014). APWG[1] (Anti Phishing Working Group) defines phishing as the act of stealing personal sensitive information such as usernames, passwords, credit card information and online banking details using either social engineering techniques or technical subterfuge. The social engineering aspect of phishing sends electronic messages (i.e. email, social media, text, online advertisements) to lure users to enter their user-credentials at a spoofed web page which looks and feels identical to a legitimate site (Zhang et al., 2007). On the other hand, it could also install a malicious software application on the user's device or to just open an attachment. For example, a victim might get the same look and feel email from his bank (with the bank logo and colours), asking him to click on a link given in the email body to visit the bank's website and verify the account details for a given security reason. The link will be directed to a counterfeit website of the targeted bank. If the user gets deceived by the similarity of the fake website's user interface, s/he will then enter the sensitive information even without noticing that s/he has become a victim of phishing. Phishing attacks use social engineering techniques (Khonji et al., 2011) as it's easy to take advantage of the human mentality by misusing how humans assign meaning to content (i.e. phishing is also known as semantic attack (Arachchilage et al., 2016)) rather than breaking into the systems straightaway (Kumaraguru et al., 2007).

As of 2019 Webroot threat report[2], the number of phishing sites detected has grown to 220% between January and December of 2018. According to the APGW 2019 trend report, the number of unique phishing websites reported in the month of March 2019 is 81122. As of 2018 FBI (Federal Bureau of Investigation) Internet Crime Complaint Centre (IC3) report[3], more than $1.3 billion financial losses have been reported in 2018 under phishing based cyber-attacks (including phishing-based Business Email Compromise (BEC) and E-mail Account Compromise (EAC) attacks). Apart from financial loss, targeted phishing attacks (spear phishing) can damage the reputation of organisations and public figures (Gupta et al., 2018) by exposing their private details to the public. John Podesta, the chairman of Hillary Clinton's presidential campaign received an email on the 19 of March 2016, with a password reset link, stating that his Gmail account has been compromised[4] and in need of immediate password rest. Induced by fear, Podesta disclosed his Gmail login credentials at the provided counterfeit website of the Gmail's password reset page. Attackers got access to all important work emails including inner details of the Clinton campaign, internal political conflicts of the Clinton Foundation and even Podesta's iCloud account password, which resulted in changing the public opinion on Hillary Clinton and her association's professional conduct.

Anti-phishing educational interventions have a major role in combating against phishing attacks by transforming "human" the weakest link in cybersecurity (Arachchilage et al., 2016), to the strongest defence. Despite the number of phishing awareness mechanisms developed over the last few decades, the phishing threat is growing (APWG, 2019). Previous research in anti-phishing education have focused on improving teaching methods (i.e. game based anti-phishing teaching (Arachchilage and Love, 2014)), user interactions (Dixon et al., 2019) and user behaviour modification (Arachchilage et al., 2016; Downs et al., 2007) as the means of mitigating phishing threat. User behaviour modifications to training users to pay attention to the URL (Uniform Resource Locator) of the website is one of the key points in every anti-phishing program (Arachchilage, 2012; Kirlappos and Sasse, 2012). Alsharnouby et al. (2005) conducted a research study using an eye-tracking device which revealed that users are paying attention to the URLs as they were trained in anti-phishing awareness programs.

However, even though users pay significant attention to the URL, majority of ordinary users fail to recognise contemporary phishing URLs (Alsharnouby et al., 2015) if the particular obfuscation technique was not covered in training programs. Therefore, one can argue that existing anti-phishing educational interventions are limited in success due to their outdated teaching content incorporated. Moreover, teaching outdated anti-phishing techniques might not help combat contemporary phishing attacks. This research focuses on investigating the obfuscation techniques of phishing URLs used in anti-phishing education against the contemporary phishing attacks reported in PhishTank.com.

---

[1] http://docs.apwg.org/reports/apwg\_trends\_report\_q1\_2019.pdf
[2] https://www-cdn.webroot.com/9315/5113/6179/2019_Webroot_Threat_Report_US_Online.pdf
[3] https://www.ic3.gov/media/annualreport/2018\_IC3Report.pdf
[4] https://www.cbsnews.com/news/the-phishing-email-that-hacked-the-account-of-john-podesta

Section 2 of the paper discusses the related work and section 3 describes the methodology and design followed in the research. Section 4 presents the results obtained by following various analysis according to the research design and discuss what's missing in the phishing awareness mechanisms that are required to fight against the contemporary phishing threats. Finally, section 5 provides the conclusion of the research and possible future work to follow-up the findings in this paper.

## 2　Related Work

The approaches to mitigate phishing threat takes two directions (Qabajeh et al., 2018). One is the development of anti-phishing tools and security mechanisms to automate phishing threat prevention within the computer systems, servers and networks. The other is the development of anti-phishing awareness programs to prevent threats at the end-user level (Alsharnouby et al., 2015). Automated anti-phishing tools are not foolproof (Wen et al., 2019) as they miss out at least 20 per cent of phishing attacks (Almomani et al., 2013). And particularly until a phishing URL is detected and registered in a spam database (Heartfield and Loukas, 2018), it's disastrous to the public. The time taken to detect a new phishing attack depends on its complexity but estimated as 24 to 96 hours. However, the Phishing Threat Report by Verizon[5] confirms that most people who are going to click on a malicious link in a phishing email do so in just over an hour and that first victim of a phishing attack falls for it within 82 seconds of the encounter. Adding to that, more than 1.5 million new phishing sites are getting created every month[6]. Therefore, it's a far too long goal for automated anti-phishing tools to detect all new phishing attack as soon as they appear and protect the early victims. On the contrary, phishing awareness programs gives a little hope as they can fight to phish at the end-user's level including zero-day attacks.

Research in phishing awareness takes different paths such as educational programs on using security alert tools and indicators such as Google Safe Browsing[7], IsThisLegit[8] (Herzberg and Amir, 2009), training programs with simulated phishing attacks such as knowbe4[9], Cofense[10], shearwater[11], anti-phishing games for end-users (Arachchilage et al., 2016; Sheng et al., 2008; Wen et al., 2019), behavioural studies to understand human weakness against phishing (Dhamija et al., 2006; Downs et al., 2007), and framework development for phishing threat measurement and awareness (Aleroud and Zhou, 2017; Garera et al., 2007).

In developing a phishing awareness program, apart from identifying efficient methods of teaching, selecting the content of the program is vital. As mentioned earlier, training users to examine the URL of a website is one of the key areas in anti-phishing education. Phishing awareness programs and anti-phishing games (Arachchilage and Love, 2014, 2016; Sheng et al., 2008) have given priority in training users to identify phishing URLs using the cues embodied within the URL itself. The nature of phishing attacks and obfuscating techniques used in phishing URLs, have been analysed and discussed over a decade ago (Dhamija et al., 2006; Garera et al., 2007). The malicious URL features identified during machine learning attempts on phishing URL classification research (Garera et al., 2007; Khonji et al., 2011; Ma et al., 2009; Whittaker et al., 2010) reveals more information on phishing attack demographics. Anti-phishing educational programs developed in the past have used these classification results that are visible to the human eye (Alsharnouby et al., 2015; Arachchilage et al., 2016).

### 2.1　URL obfuscating Techniques

The term URL obfuscation stands for the innovative ways used by phishers to create deceiving phishing URLs by carefully placing cue words in a URL to trick human mind. The interesting study by Garera et al. (Garera et al., 2007) has categorised prominent obfuscation techniques by examining the blacklist of phishing URLs maintained by Google in 2007 (Garera et al., 2007).

---

[5] https://enterprise.verizon.com/resources/reports/DBIR_2018_Report_execsummary.pdf
[6] https://www-cdn.webroot.com/9315/5113/6179/2019_Webroot_Threat_Report_US_Online.pdf
[7] https://safebrowsing.google.com
[8] https://github.com/duo-labs/isthislegit
[9] https://www.knowbe4.com
[10] https://cofense.com
[11] https://www.shearwater.com.au/

- Type I: Obfuscating the Hostname with an IP Address - URL's hostname (Netloc component) is obfuscated by the phisher using an IP address. Very often the IP address is also represented in hex or decimal format rather than the common dotted quad form. An example of this type of URL is: http://67.210.122.222/ apple/login.

- Type II: Obfuscating the host with another domain - A valid looking domain name is used in the "path" component of the URL and trick the user to see it as a redirect URL. https://recovery-confrim-paqe.cf/?facebook.com=chekpoint.

- Type III: Obfuscating with large hostnames - In this type of obfuscation, phisher appends a large string of words at the end of a genuine-looking domain name to deceive the target user.

- Type IV: Domain name unknown or misspelt - The phishing URL may contain a domain name which is different to the target organisation. This can either be an unknown domain or a misspelt version of the legitimate domain. As example is: http://www.g00g1e.com.

There has been much research in the past actively using these obfuscation categories with different features in phishing URL identification, yet have not introduced new categories to the literature (Cui et al., 2018; Darling et al., 2015; Le et al., 2011; Musuva et al., 2019; Qabajeh et al., 2018; Shirazi et al., 2018). Especially, the phishing awareness mechanisms developed since then have used these URL obfuscation techniques as the basis in organising the teaching content (Arachchilage et al., 2016; Sheng et al., 2008, 2008; Wen et al., 2019). The content of anti-phishing educational games also follows the same pattern of simulated phishing attacks where they synthetically develop the possible phishing URLs by following the URL obfuscation techniques. Furthermore, Le et al. in 2011 have revealed that URL features alone are enough to identify malicious URLs (Le et al., 2011). Within the URL obfuscation techniques, the Bag of Word approach has also been employed where users are trained to suspect URLs that contain top trick words identified. Simulated phishing awareness programs[10] [11] highly use these words to train people to look carefully.

Since the URL obfuscation techniques presented by Garera et al. 2007 have remained unchanged for many years now, the topicality of these techniques against the contemporary phishing URLs is an important research question. Importantly the anti-phishing tools and awareness programs developed based on these obfuscation techniques run the risk of being outdated (Arachchialge, 2012). It's interesting to see that, while developing PhishDef (Le et al., 2011) in 2011 the researchers have extended the word-based features by Garera et al. (2007), yet have used all other concepts without any change. We identified that existing phishing awareness interventions haven't conducted a thorough analysis of contemporary phishing URLs before program development as they are depending on the phishing URL categorisations of Garera et al. 2007.

On the other hand, the evaluation criteria employed in existing phishing awareness programs are also focused on measuring user interaction improvement and success rate of user behaviour modification techniques with pre and post-tests (Arachchilage et al., 2016; Kirlappos and Sasse, 2012; Sheng et al., 2008; Wen et al., 2019; Wu, 2006). However, the primary expectation of anti-phishing education is to improve users' security knowledge to combat against the contemporary phishing attacks (Musuva et al., 2019). Cybercriminals are continuously learning and keep inventing various new strategies (Kumaraguru et al., 2007). Therefore, investigating the URL obfuscation techniques used in anti-phishing education against the contemporary phishing attacks is a timely requirement.

## 3 Methodology and Research Design

This research focused on investigating the obfuscation techniques of phishing URLs used in anti-phishing education against the contemporary phishing attacks. Previous research on phishing classification research has been investigated in the binary classification of phishing attacks where they decided whether or not a given URL is malicious (Darling et al., 2015; Whittaker et al., 2010). On the other hand, phishing attack measurement studies such as Garera et al. 2007 have categorised the phishing attacks based on their characteristics. Therefore, we conducted a follow-up analysis of the URL obfuscation techniques against contemporary phishing URLs from PhishTank.com[12].

### 3.1 Data Used

Our study obtained a contemporary phishing URL dataset downloaded from PhishTank[12] on 05th May 2019. PhishTank.com[12] is a well-known non-profit community database that maintains a blacklist of phishing URLs reported and verified by volunteers. For this study, we downloaded the dataset in a CSV file and it contained a total of 10078 verified phishing URLs. The collected phishing URLs ranged from real-world phishing attempts reported between 2008 and 2019. We used the Python programming language[13] to conduct the initial feature analysis of phishing URLs due to the cost-effective and supportive nature. URL Parsing Library (URLLib[14]) was used for the lexical analysis on the URL components described in Figure1.

---

[12] https://www.phishtank.com

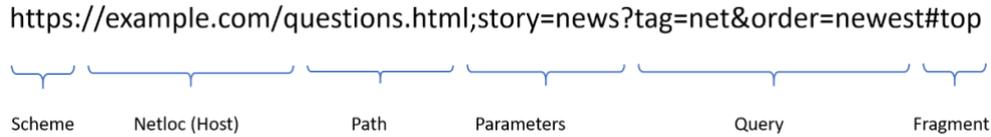

Figure 1: URL structure

## 3.2 Structure of URLs

Uniform Resource Locator (URL), or the web address, is a reference to a web resource that specifies its location on a computer network (Berners-Lee et al., 2005). As highlighted in the literature, previous research has proved that careful attention to URL lexical is sufficient enough (Le et al., 2011) to identify the legitimacy of a website. IETF (Internet Engineering Task Force) (Berners-Lee et al., 2005) has specified the URL structure in 6 components as per Figure 1. Table 1 shows the URL component distribution in our dataset. We analysed the PhishTank[12] dataset against below URL components to identify the URL obfuscation techniques that can be used to train end-users to detect phishing threats.

| URL Component | Description | % of dataset (2019) (N=10078) |
|---|---|---|
| Scheme | network communication protocol i.e. http, "https", "ftp" "mailto" | 100% |
| Netloc | network location - primary hostname or domain | 100% |
| Path | hierarchical path from root domain to the target web page | 99.5% |
| Parameters | URL parameter (Path Param) to identify a specific resource | 0.4% |
| Query | provides filtration to resources available in the given path | 19.8% |
| Fragment | contains a fragment identifier | 0.7% |

*Table 1. URL components distribution in our dataset*

## 3.3 Feature Analysis

This research followed a similar approach of phishing URL analysis presented in the work by Garera et al. 2007. Their study has categorised website features into four groups as Page Based, Domain Based, Type Based and Word-Based features. From that Type Based features (URL obfuscation types) and Word-Based features (suggestive token words found in URLs) have been analysed under website URL features. Following that, in this research, the contemporary phishing URLs obtained from PhishTank[11] were analysed under Type-based features and Word-based features and compared with the results presented in the study by Garera et al. (2007).

Our study retrieved 10078 phishing URLs from PhishTank[11], whereas Garera et al. (2007) have had only 1245 phishing URLs in their training blacklist dataset from Google. The results were compared in terms of percentages to better understand the composition of URL obfuscation techniques. Under type based features we analysed the availability of Type I (Obfuscating hostname with an IP Address) and Type II (Obfuscating the host with another domain) within the contemporary phishing URL dataset and compared that with the results of Garera et al. (Garera et al., 2007). Type III obfuscation technique (Obfuscating with large URLs) is on the length of the URL, especially on the misleading characters added at the end of a legitimate domain. Then the study was conducted on the word-based features presented in Garera et al. (2007) to understand the suggestive word tokens in contemporary phishing URLs.

---

[13] https://www.python.org
[14] https://docs.python.org/3/library/urllib.html

## 4　Results and Discussion

We investigated the contemporary phishing attacks reported in real-world against the URL obfuscation techniques that are used in anti-phishing educational content development. Our results showed that the composition of URL obfuscation techniques has significantly changed over time. The Type I (Obfuscating hostname with an IP Address), which was a popular obfuscation technique in 2007 with 63.6% of occurrence only has an insignificant value of 1.03% in our contemporary dataset. The usage of Type II obfuscation technique (Obfuscating the host with another domain) also have reduced in the contemporary dataset by having 26.9% of occurrence.

There were many phishing URLs in our contemporary dataset that do not fit into any of the four existing URL obfuscation techniques (i.e. four categories identified by Garera et al. (2007)). Therefore, we analysed the contemporary phishing URL dataset itself to identify new obfuscation techniques. We revealed, in addition to Garera et al's. (2007) 4 categories, two major obfuscation techniques which have not been incorporated into anti-phishing teaching content. It also showed that the most common word tokens used by phishers have changed over time. Therefore, we further analysed the contemporary dataset from PhishTank[15] to identify new suggestive word tokens.

### 4.1　Type Based Features

Table 2 shows the comparison of Type I (Obfuscating hostname with an IP Address i.e. http:// 67.210.122.222/apple/login) and Type II (Obfuscating the host with another domain i.e. https:// recovery-confrim-paqe.cf/?facebook.com=chekpoint) URL obfuscation techniques. As shown in Table 2, Type I obfuscation techniques is rarely found in contemporary phishing URLs, we obtained in 2019. Type II obfuscation technique has also decreased by 12.8% in 2019. Data available in PhishTank.com[12] under the spoofed organisation (which indicates the name of the organisation used in the URL path) was used to identify the Type II Obfuscation Technique. In our dataset, there were only 2718 URLs with the identified spoofed organisation. However, the majority of the URLs had unresolved target organisations due to the complexity of the URL formulation. For example, "https://www.wireless-handsets.com/blueprintcsg/verificationAttempt.php?sf58g…" seems like a wireless headset sales company and users fell for it. However, there is no existing company or legitimate website under that name and this falls under the category of unresolved target organisations.

| URL Obfuscation Technique | Garera et al. (2007) (N=1245) | Current dataset (2019) (N=10078) |
|---|---|---|
| Type I - Obfuscating hostname with an IP Address | 63.6% | 1.03% |
| Type II - Obfuscating the host with another domain | 39.7% | 26.9 % |

*Table 2. Type I and Type II comparison on phishing URLs from 2007 (Garera et al., 2007) vs 2019*

In analysing the Type III obfuscation technique (Obfuscating with large URLs), the URLs with large domain names were identified and then we checked if that domain has words placed intentionally (delimited by '-', '_', '=', '?', '%') to imitate a target organisation name. For example, "https://nz1webapps7mpp3manage-my-papl-account.felixkot.biz/signin…", has a large domain name with suggestive words placed in the subdomain to make it look like it is linked to paypal.com. In the analysis by Garera et al. (2007), the average length of additional characters appended to the domain name of phishing URL was 7.34 characters and the maximum length was 63. In our phishing dataset, the average length of additional characters appended to the domain was 9.32 with the maximum length of 73.88 characters. With the results, it seems the Type III obfuscation techniques apply to current data. In early 2019, there has been a few attempts of phishing with more than 1000 characters in the URL (Abrams, 2019) using the Obfuscation with a large URL technique (Type III). The largest URL in our dataset had 1149 characters. Therefore, we can expect phishing attacks with more complex URLs in the future.

### 4.2　New URL Obfuscation Techniques for Anti-Phishing Education

As our findings revealed that URL obfuscation techniques identified by Garera et al. 2007 (that have been incorporated into existing anti-phishing teaching content) are not aligning with the real-world contemporary phishing threats, we analysed the contemporary dataset itself to investigate new URL obfuscation techniques. The results below would look obvious for any technically savvy individual, yet not included in anti-phishing teaching content, and therefore non-technical end-users are highly susceptible.

---

[15] https://www.phishtank.com

### 4.2.1 Type V: Obfuscating with HTTPS Schema

Technically the HTTPS schema represents the secure and encrypted data communication protocol between the user's browser and the target site. Browsers display a green padlock icon in the address bar only to represent secure communication and it is not related to the legitimacy of the website. However according to the survey conducted by PhishLab[16] more than 80% of non-technical users believe green lock and https indicate that a website is either legitimate and/or safe[16].

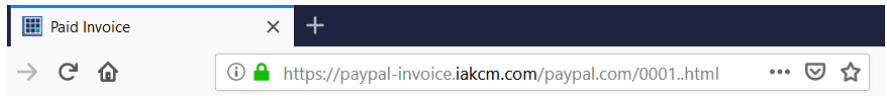

*Figure 2: Https with Green Padlock for phishing URL that targets Paypal.com*

Figure 2 shows a phishing website with SSL/TLS certificate and green padlock. Our findings demonstrate that phishers are increasingly exploiting user's misbelief in https schema for their advantage. Table 6 displays the 3231 https URLs we found in our dataset with increased usage pattern over the years.

| Year | Occurrences of https | Total phishing attacks of the year in the current dataset | Percentage of https URLs |
|---|---|---|---|
| 2019 (until 05/05/2019) | 1477 | 3930 | 37.58% |
| 2018 | 1543 | 4068 | 37.93% |
| 2017 | 182 | 1249 | 14.57% |
| 2016 | 16 | 488 | 3.27% |
| 2015 | 3 | 233 | 1.28% |

*Table 6. Https protocol usage in Phishing URLs*

Here we introduce the Type V obfuscation technique – "Obfuscating with HTTPS schema", as an additional category type to phishing URL obfuscation categories proposed by Garera et al. 2007.

### 4.2.2 Type VI: Obfuscating with Internationalized Domain Names

Internationalized Domain Names (IDNs) are domain names represented by non-ASCII local language characters (Richard Ishida (W3C), 2008). IDN was the solution to improve the usability of the Internet for millions of non-English speaking users by enabling to have domain names in non-ASCII characters. Punycode is an algorithm employed by web protocols to transform a Unicode string into an ASCII string (RFC3492[17]). Punycode sequences always start with the prefix "xn--" simply to distinguish IDNs from ordinary domain names. The punycode for the URL in Figure 3 is "xn--oy2b35ckwhba574atvuzkc.com".

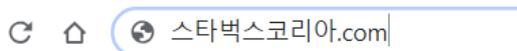

*Figure 3: IDN with local language*

However, IDN has been exploited by phishers (Liu et al., 2018) in their advantage to mislead the end-users with Unicode characters that have similar appearances to ASCII characters, which are undetectable in browser address bars. For example, "https://www.apple.com" is not the actual apple URL as the "a" of apple.com is a Cyrillic "a" (U+0430) rather than the ASCII character "a" (U+0061). The punycode for the Cyrillic character URL is https://www.xn--pple-43d.com/. We found 24 IDN URLs in our dataset with increased usage pattern over the years in Table 7.

| Year | Occurrences in Current dataset (2019) |
|---|---|
| 2019 (until 05/05/2019) | 12 |
| 2018 | 8 |

---

[16] https://info.phishlabs.com/blog/quarter-phishing-attacks-hosted-https-domains
[17] https://www.ietf.org/rfc/rfc3492.txt

| | |
|---|---|
| 2017 | 1 |
| 2016 | 1 |
| 2015 | 0 |

*Table 7. Usage of IDN in Phishing URLs for camouflage*

As the usage of IDN in 2019 until 05/05/2019 has already passed the total number of usages before 2019, we can expect more IDN based phishing attacks in the coming years. Therefore, we introduce this as the Type VI obfuscation technique – "Obfuscating with Internationalized Domain names", as an extension to phishing URL obfuscation types proposed by Garera et al (2007).

### 4.3  Word Based Features

Word based feature analysis is a commonly used technique in most of automated phishing detection tools (Cui et al., 2018; Jain and Gupta, 2018; Khonji et al., 2011; Marchal et al., 2016) as it provides more contextual information on the legitimacy of the URL text. Also, the blacklisted bag of words enhances the detection of Type IV URL obfuscation technique (obfuscating with unknown or misspelt Domain name). As per the findings of Garera et al (2007), phishing URLs are found to be containing several suggestive word tokens frequently. We examined our contemporary phishing URL dataset against these words to understand whether or not they are still prominent. The comparisons were non case-sensitive and Boolean (where we only checked whether or not a given word is present in a URL).

| Special Words | Garera et al. (2007) (N=1245) | Current dataset (2019) (N=10078) |
|---|---|---|
| confirm | 4.25% | 0.893% |
| account | 4.9% | 5.040 % |
| banking | 7.95% | 0.188% |
| secure | 9.88% | 3.006% |
| ebayisapi | 13.9% | 0.019% |
| webscr | 14.2% | 0.367% |
| login | 21.53% | 13.177% |
| signin | 23.29% | 2.937% |

*Table 3. Word Token feature comparison on phishing URLs in 2007 (Garera et al., 2007) and current dataset from PhishTank[12] (2019)*

From the results in Table 3, we can see that except the words "account" and "login" other special words identified by Garera et al. (2007) are no longer found common in the contemporary phishing URLs. The usage of word "signin" as dropped by more than 20%. The decreased numbers in words like "webscr" or "ebayisapi" suggests that the changes in the web development technologies might have influenced phishers and the way they phish.

In developing PhishDef, the anti-phishing tool (Le et al., 2011), researchers have added four more words ("paypal", "free", "lucky", "bonus") to the suggestive word list of Garera et al. 2007 as an extension. Therefore, we analysed the frequency of these four words in our dataset to understand the current situation of using words based features on phishing URLs in 2019. Results in Table 4 showed that the frequency of the words also has deprived now.

| Special Words | Percentage in Current dataset (2019) (N=10078) |
|---|---|
| paypal | 2.431% |
| free | 1.716% |
| lucky | 0.039% |
| bonus | 0.089% |

*Table 4. Usage of Word Token features added by Le et al. (Le et al., 2011) current dataset from PhishTank[12] (2019)*

#### 4.3.1  Lexicon for Phishing URLs in 2019

The results we discussed in above revealed that frequent suggestive words identified by previous research do not occur in contemporary phishing attacks, we further examined the word frequency in the dataset retrieved from PhishTank[11]. Table 5 shows the ten most common words in our contemporary phishing URL dataset and apart from "login" and "account" all other 8 are new lexicons. Here we followed the same filtration criteria used by Garera et al. 2007 and discarded all tokens with length<5 as they contained several common technical terms used in website development.

| Words | Total No. of URLs | Percentage in Current dataset (2019) (N=10078) |
|---|---|---|
| Login | 1328 | 13.177% |
| Account | 508 | 5.040% |
| Content | 505 | 5.010% |
| Include | 496 | 4.921% |
| Online | 478 | 4.743% |
| Sites | 461 | 4.574% |
| Admin | 455 | 4.514% |
| Email | 431 | 4.276% |
| Secur | 422 | 4.187% |
| image | 394 | 3.909% |
| update | 304 | 3.016% |

*Table 5. Frequency of Word Tokens in the current dataset from PhishTank[12] (2019)*

# 5 Conclusion and Future Work

This research focused on investigating the phishing URL obfuscation techniques incorporated in anti-phishing education against the contemporary phishing attacks. The URL obfuscation techniques presented a decade ago by Garera et al. 2007 are still being used in anti-phishing educational interventions. Over time attackers have abandoned the already identified and educated methods of URL obfuscation techniques and keep inventing new ways of deceiving. Our research findings revealed that the URL obfuscation techniques used in existing anti-phishing teaching do not help people to detect contemporary phishing attacks. The Type I obfuscation technique (Obfuscating hostname with an IP Address) presented by Garera et al. 2007 had 33.32% representation of the total phishing attacks, yet in our contemporary dataset, it has only 1.03% of insignificant usage. Adding to that, Type II (Obfuscating the host with another domain) technique has also decreased by 12.8%. Therefore, one can argue that existing anti-phishing educational interventions are limited in success in mitigating contemporary phishing attacks due to their outdated teaching content incorporated. Moreover, teaching outdated anti-phishing techniques might not help combating contemporary phishing attacks. In addition to Garera et al's (2007) categorisation, we have also identified two new URL obfuscating techniques, which are:

1. Type V - Obfuscating with HTTPS schema.
2. Type VI - Obfuscating with Internationalized Domain names

Identified two new obfuscation techniques have a trending usage for phishing where 37.58% of phishing URLs of 2019 (until 05/05/2019) have used https schema. 12 IDN URLs have been reported until 05/05/2019, which is higher than the total number of IDNs reported in previous years. These new obfuscation techniques need to be incorporated into contemporary teaching content (in our case, phishing URLs) of anti-phishing educational interventions to better educate end users to prevent phishing attacks. Apart from that, these two new URL obfuscation techniques can also be used as features in creating new automated phishing detection tools. In future work, we will analyse various contemporary phishing datasets to identify more hidden URL obfuscation techniques, which have not been incorporated in developing anti-phishing educational interventions.

# 6 References

Abrams, L., 2019. Weird Phishing Campaign Uses Links with Almost 1,000 Characters https://www.bleepingcomputer.com/news/security/weird-phishing-campaign-uses-links-with-almost-1-000-characters/.

Aleroud, A., Zhou, L., 2017. Phishing environments, techniques, and countermeasures: A survey. Computers & Security 68, 160–196. https://doi.org/10.1016/j.cose.2017.04.006

Almomani, A., Gupta, B.B., Atawneh, S., Meulenberg, A., Almomani, E., 2013. A survey of phishing email filtering techniques. IEEE Communications Surveys and Tutorials 15, 2070–2090. https://doi.org/10.1109/SURV.2013.030713.00020

Alsharnouby, M., Alaca, F., Chiasson, S., 2015. Why phishing still works: user strategies for combating phishing attacks. International Journal of Human-Computer Studies.

APWG, 2019. Phishing Activity Trends: 1st Quarter 2019. APWG.

Arachchilage, N.A.G., Love, S., 2014. Security awareness of computer users: A phishing threat avoidance perspective. Computers in Human Behavior 38, 304–312. https://doi.org/10.1016/j.chb.2014.05.046

Arachchilage, N.A.G., Love, S., Beznosov, K., 2016. Phishing threat avoidance behaviour: An empirical investigation. Computers in Human Behavior 60, 185–197. https://doi.org/10.1016/j.chb.2016.02.065

Arachchilage, N.A.G., 2012. Security awareness of computer users: A game based learning approach. School of Information Systems ,Computing and Mathematics ,Brunel University.

Berners-Lee, T., Fielding, R., Masinter, L., 2005. Uniform Resource Identifier (URI): Generic Syntax. RFC 3986. https://doi.org/10.17487/rfc3986

Chen, S.-Y., Jeng, T.-H., Huang, C.-C., Chen, C.-C., Chou, K.-S., 2017. Doctrina, in: Proceedings of the 2017 the 7th International Conference on Communication and Network Security - ICCNS 2017. ACM Press, New York, New York, USA, pp. 67–75. https://doi.org/10.1145/3163058.3163061

Cui, B., He, S., Yao, X., Shi, P., 2018. Malicious URL detection with feature extraction based on machine learning. International Journal of High Performance Computing and Networking 12, 166. https://doi.org/10.1504/ijhpcn.2018.094367

Darling, M., Heileman, G., Gressel, G., Ashok, A., Poornachandran, P., 2015. A lexical approach for classifying malicious URLs, in: 2015 International Conference on High Performance Computing & Simulation (HPCS). IEEE, pp. 195–202. https://doi.org/10.1109/HPCSim.2015.7237040

Dhamija, R., Tygar, J.D., Hearst, M., 2006. Why phishing works, in: Proceedings of the SIGCHI Conference on Human Factors in Computing Systems - CHI '06. ACM Press, New York, New York, USA, p. 581. https://doi.org/10.1145/1124772.1124861

Dixon, M., Gamagedara Arachchilage, N.A., Nicholson, J., 2019. Engaging Users with Educational Games: The Case of Phishing, in: Extended Abstracts of the 2019 CHI Conference on Human Factors in Computing Systems, CHI EA '19. ACM, New York, NY, USA, p. LBW0265:1–LBW0265:6. https://doi.org/10.1145/3290607.3313026

Downs, J.S., Holbrook, M., Cranor, L.F., 2007. Behavioral response to phishing risk, in: Proceedings of the Anti-Phishing Working Groups 2nd Annual ECrime Researchers Summit on - ECrime '07. ACM Press, New York, New York, USA, pp. 37–44. https://doi.org/10.1145/1299015.1299019

Garera, S., Provos, N., Chew, M., Rubin, A.D., 2007. A framework for detection and measurement of phishing attacks, in: ACM Workshop on Recurring Malcode (WORM '07). p. 1. https://doi.org/10.1145/1314389.1314391

Gupta, B.B., Arachchilage, N.A., Psannis, K.E., 2018. Defending Against Phishing Attacks: Taxonomy of Methods, Current Issues and Future Directions. Telecommun. Syst. 67, 247–267. https://doi.org/10.1007/s11235-017-0334-z

Heartfield, R., Loukas, G., 2018. Detecting semantic social engineering attacks with the weakest link: Implementation and empirical evaluation of a human-as-a-security-sensor framework. Computers & Security 76, 101–127. https://doi.org/10.1016/j.cose.2018.02.020

Herzberg, A., Amir, 2009. Why Johnny can't surf (safely)? Attacks and defenses for web users. Computers & Security 28, 63–71. https://doi.org/10.1016/j.cose.2008.09.007

Jain, A.K., Gupta, B.B., 2018. PHISH-SAFE: URL features-based phishing detection system using machine learning, in: Advances in Intelligent Systems and Computing. Springer, Singapore, pp. 467–474. https://doi.org/10.1007/978-981-10-8536-9_44

Khonji, M., Iraqi, Y., Jones, A., 2011. Lexical URL analysis for discriminating phishing and legitimate websites, in: Proceedings of the 8th Annual Collaboration, Electronic Messaging, Anti-Abuse and Spam Conference on - CEAS '11. ACM Press, New York, New York, USA, pp. 109–115. https://doi.org/10.1145/2030376.2030389

Kirlappos, I., Sasse, M.A., 2012. Security education against Phishing: A modest proposal for a Major Rethink. IEEE Security and Privacy 10, 24–32. https://doi.org/10.1109/MSP.2011.179

Kumaraguru, P., Rhee, Y., Acquisti, A., Cranor, L., Hong, J., Nunge, E., 2007. Protecting people from phishing: The design and evaluation of an embedded training email system, in: Conference on Human Factors in Computing Systems - Proceedings. pp. 905–914. https://doi.org/10.1145/1240624.1240760

Le, A., Markopoulou, A., Faloutsos, M., 2011. PhishDef: URL names say it all, in: 2011 Proceedings IEEE INFOCOM. IEEE, pp. 191–195. https://doi.org/10.1109/INFCOM.2011.5934995

Liu, B., Lu, C., Li, Z., Liu, Y., Duan, H., Hao, S., Zhang, Z., 2018. A Reexamination of Internationalized Domain Names: The Good, the Bad and the Ugly, in: 2018 48th Annual IEEE/IFIP International Conference on Dependable Systems and Networks (DSN). IEEE, pp. 654–665. https://doi.org/10.1109/DSN.2018.00072

Ma, J., Saul, L.K., Savage, S., Voelker, G.M., 2009. Beyond blacklists: learning to detect malicious web sites from suspicious URLs, in: Proceedings of the 15th ACM SIGKDD International Conference on Knowledge Discovery and Data Mining - KDD '09. ACM Press, New York, New York, USA, p. 1245. https://doi.org/10.1145/1557019.1557153

Marchal, S., Saari, K., Singh, N., Asokan, N., 2016. Know Your Phish: Novel Techniques for Detecting Phishing Sites and Their Targets, in: Proceedings - International Conference on Distributed Computing Systems. IEEE, pp. 323–333. https://doi.org/10.1109/ICDCS.2016.10

Musuva, P.M.W., Getao, K.W., Chepken, C.K., 2019. A new approach to modelling the effects of cognitive processing and threat detection on phishing susceptibility. Computers in Human Behavior 94, 154–175. https://doi.org/10.1016/j.chb.2018.12.036

Qabajeh, I., Thabtah, F., Chiclana, F., 2018. A recent review of conventional vs. automated cybersecurity anti-phishing techniques. Elsevier. https://doi.org/10.1016/j.cosrev.2018.05.003

Richard Ishida (W3C), 2008. An Introduction to Multilingual Web Addresses.

Sheng, S., Magnien, B., Kumaraguru, P., Acquisti, A., Cranor, L.F., Hong, J., Nunge, E., 2008. Anti-Phishing Phil, in: Proceedings of the 3rd Symposium on Usable Privacy and Security - SOUPS '07. ACM Press, New York, New York, USA, p. 88. https://doi.org/10.1145/1280680.1280692

Shirazi, H., Bezawada, B., Ray, I., 2018. Kn0w Thy Doma1n Name, in: Proceedings of the 23nd ACM on Symposium on Access Control Models and Technologies - SACMAT '18. ACM Press, New York, New York, USA, pp. 69–75. https://doi.org/10.1145/3205977.3205992

Wen, Z.A., Lin, Z., Chen, R., Andersen, E., 2019. What.Hack: Engaging Anti-Phishing Training Through a Role-playing Phishing Simulation Game, in: Proceedings of the 2019 CHI Conference on Human Factors in Computing Systems - CHI '19. ACM Press, New York, New York, USA, pp. 1–12. https://doi.org/10.1145/3290605.3300338

Whittaker, C., Ryner, B., Nazif, M., 2010. Large-Scale Automatic Classification of Phishing Pages., in: NDSS '10.

Wu, M., 2006. Fighting Phishing at the User Interface. Massachusetts Institute of Technology.

Zhang, Y., Zhang, Y., Egelman, S., Cranor, L., Hong, J., 2007. Phinding phish: Evaluating anti-phishing tools, in: In Proceedings of the 14th Annual Network & Distributed System Security Symposium (NDSS 2007). San Diego, CA.

## Copyright